# Role Classification of Hosts within Enterprise Networks Based on Connection Patterns

Godfrey Tan
*MIT*
*godfreyt@mit.edu*

Massimiliano Poletto
*Mazu Networks*
*maxp@mazunetworks.com*

John Guttag and Frans Kaashoek
*MIT*
{*guttag,kaashoek*}*@mit.edu*

## Abstract

Role classification involves grouping hosts into related roles. It exposes the logical structure of a network, simplifies network management tasks such as policy checking and network segmentation, and can be used to improve the accuracy of network monitoring and analysis algorithms such as intrusion detection.

This paper defines the role classification problem and introduces two practical algorithms that group hosts based on observed connection patterns while dealing with changes in these patterns over time. The algorithms have been implemented in a commercial network monitoring and analysis product for enterprise networks. Results from grouping two enterprise networks show that the number of groups identified by our algorithms can be two orders of magnitude smaller than the number of hosts and that the way our algorithms group hosts highly reflect the logical structure of the networks.

## 1 Introduction

Today, many enterprises have internal networks (intranets) that are as or more complicated than the entire Internet of a few years ago. Managing these networks is increasingly costly, and the business cost of network problems increasingly high.

Managing an enterprise network involves a number of inter-related activities, including:

**Establishing a topology.** A network's topology has a significant impact on its cost, security, and performance. An increasingly important aspect of topology design is *network segmentation*. In an effort to provide fault isolation and mitigate the spread of worms like Nimda [3] and Code Red [2], enterprises segment their networks using firewalls [4], routers, VLANs [7], and other technologies.

**Establishing policies.** Different users of a network have different privileges. Some users may have unlimited access to external networks while others may have restricted access. Some users may be limited in the amount of bandwidth they may consume, and so on. The number of policies is open-ended.

**Monitoring network performance.** Almost every complex network suffers from various localized performance problems. Network managers must detect these problems and take action to correct them.

**Detecting and responding to security violations.** Increasingly, networks are coming under attack. Sometimes the targets are chosen at random, as in most virus-based attacks, and in other cases they are picked intentionally, as with most denial-of-service attacks. These attacks often involve compromised computers within the enterprise network. Early detection of attacks plays a critical role in reducing the damage.

Conducting these activities on a host-by-host basis is not feasible for large networks. Network managers need to extract structure from their networks so that they can think about them and make decisions at larger levels of granularity. Today, this structuring is most often done in an *ad hoc* manner that relies on administrators' best guesses about the computers, services, and users on the network. Obviously, this method has scaling problems.

This paper presents two algorithms that, used together, partition the hosts on an enterprise network into groups in a way that exposes the logical structure of a network. The *grouping algorithm* classifies hosts into groups, or "roles," based on their *connection habits*. The *correlation algorithm* correlates groups produced by different runs of the classification algorithm.

The two algorithms together provide the following properties:

1. They guarantee that a host is only grouped with other hosts that have the strongest degree of similarity in connection habits.

2. They provide a mechanism to merge groups, and give network administrators fine-grained control over the merging process, so that meaningful results can be achieved.

3. They deal with transient changes in connection patterns by analyzing the profiled data over long periods.

4. They respond to non-transient changes in connection patterns by producing a new partitioning and describing the differences between the new partitioning and the previous partitioning.

5. Their run time grows quadratically with the number of hosts in the enterprise network.

As we demonstrate in Section 6, the algorithms reduce the number of logical units that a network administrator must deal with by *one or two orders of magnitude.* The algorithms are implemented as part of an enterprise monitoring and analysis system that is in production use at several large enterprises.

Section 2 outlines the system in which the algorithms operate, and introduces an example scenario that will be used throughout the paper. Section 3 describes the models used to develop practical solutions. Section 4 and Section 5 explain the two practical algorithms for solving the role classification problem. Section 6 presents preliminary results, and Section 7 discusses related work. We conclude with discussions of our current and future work in Section 8.

## 2 System Overview

The role classification algorithms are implemented as part of a system designed to detect and respond to security violations in large enterprise networks. Such networks commonly consist of tens of thousands of computers, spread over different geographic locations. The security system consists of *probes* and a central *aggregator*. The probes analyze packets on the link or links they are attached to, and send relevant information (including IP address/port tuples) to the aggregator.

The aggregator is a scalable system that consists of one or more CPUs. It periodically runs several analysis algorithms on the data it has received from the probes. It uses the role classification algorithms to refine its analyses and to allow the administrators to describe group-based policies.

Figure 1 presents a simple enterprise network and a partitioning of computers into groups that the aggregator might produce based on the communication patterns observed by the probes. The communication patterns might indicate that hosts *Sales-1* to *Sales-N* communicate with three servers: *Mail* server, *Web* server, and *SalesDatabase* server. Similarly, the patterns might indicate that hosts *Eng-1* to *Eng-M* communicate mostly with *Mail* server, *Web* server, and *SourceRevisionControl* server.

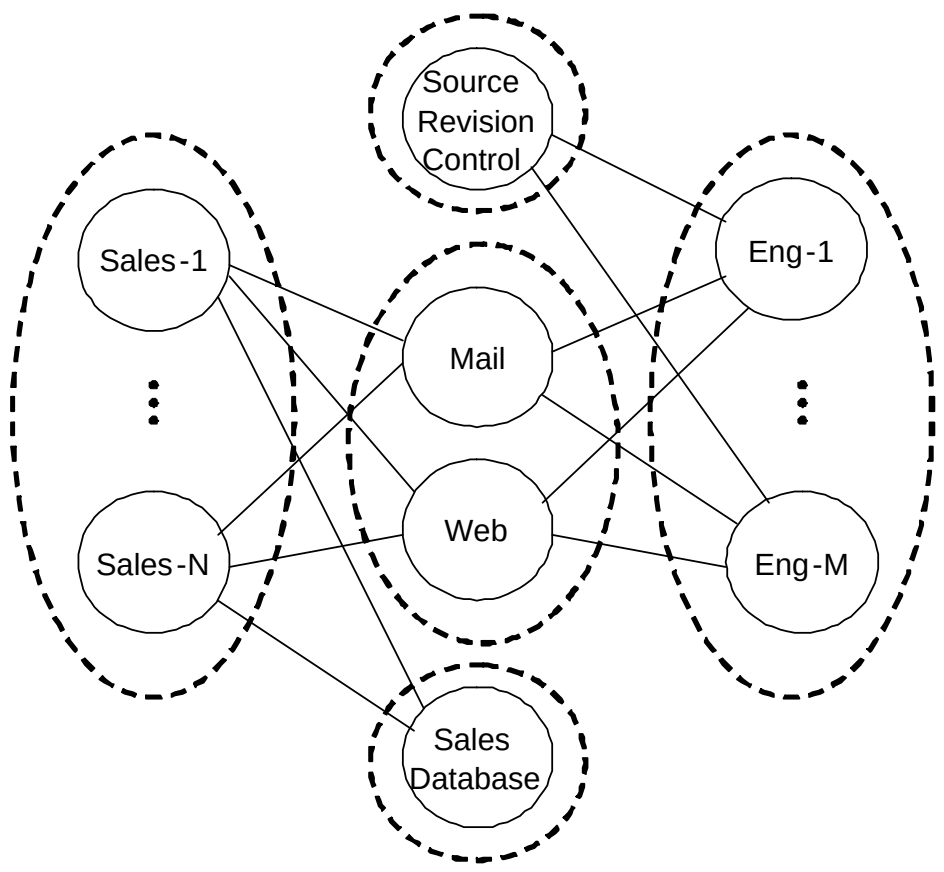

Figure 1. Grouping of related hosts based on connection patterns. Edge indicates that nodes communicate regularly. The dashed circle represents the group boundary.

Based on this information the grouping algorithm can logically divide all machines into five groups: (i) the sales group consisting of hosts *Sales-1* to *Sales-N*, (ii) the engineering group consisting of hosts *Eng-1* to *Eng-M*, (iii) the common server group consisting of *Mail* and *Web*, (iv) the sales server group consisting of *SalesDatabase* and (v) the engineering server group consisting of *SourceRevisionControl*.

The results of the grouping algorithm are currently being used in two major ways:

1. The Mazu network monitoring and detection system decides whether a host's behavior matches the expected policy setting, partly based on the history of the host's group membership. For example, if a host in the engineering group were to suddenly start opening connections to the *SalesDatabase* server, it might be a cause for alarm.

2. The network administrators review the grouping results to better understand the structure of their networks and to get useful insights for conducting network re-organization tasks such as consolidating servers and network segmentation.

The system allows a network manager to label each identified group with descriptive roles and set policies per group. The system continuously monitors the communication patterns, adjusts groups as computers come and go, flags policy violations, and raises alerts about potential security violations. Because all this information is presented on the level of groups (instead of individual hosts), a network manager is able to understand and process the changes and alerts more easily. The algorithms also provide network administrators with flexibility to control the grouping process to achieve results that highly reflect their intuitive notion of the network structure.

The algorithms presented in this paper are solely based on the connection patterns of hosts such as the set of neighboring hosts. However, the algorithms can easily be extended to use other information such as protocols and port numbers used and bytes transferred to achieve desired results. For instance, some network administrators may desire that *Mail* and *Web* servers be put in different groups. In this case, the protocol information can be used to keep the role classification algorithm from grouping together hosts that use different sets of protocols. We are currently exploring ways to expand the capability of the grouping and correlation algorithms by providing network administrators with more flexibility to achieve desired results.

The algorithms assume that the connection patterns of hosts highly reflect the logical roles that they play. For some networks where this is not true, the algorithms will not do a good job. However, we believe that hosts in a typical enterprise network that share the same logical role will demonstrate similar connection patterns.

## 3 Model

In this section, we develop a model for thinking about the grouping problem. We define the problem in the abstract, providing a model with several functions and parameters that can be adjusted to meet various goals. Later in the paper, we present and evaluate instantiations of these parameters.

- Let $I$ be the set of hosts in an enterprise network. We will use $|I|$ to denote the number of hosts in $I$.

- Let *similarity* be a commutative function from pairs of hosts in $I$ to an integer greater than or equal to $0$. Roughly speaking, if *similarity*$(h_1, h_2)$ is high, then we would like our grouping algorithm to place the hosts $h_1$ and $h_2$ in the same *group*. Defining *similarity* so that it is both efficient to compute and yields a good grouping is at the heart of the problem addressed in this paper.

- A partitioning $P$ of $I$ *respects similarity* if for all distinct groups $G_1, G_2 \in P$, $h_1, h_2 \in G_1$, and $h_3 \in G_2$,
  - *similarity*$(h_1, h_2) \geq$ *similarity*$(h_1, h_3)$
  - *similarity*$(h_1, h_2) \geq$ *similarity*$(h_2, h_3)$

We extend this definition of *similarity* to define the average similarity between a host $h_1$ and a group $G_2$, *avg_similarity*$(h_1, G_2)$, as the ratio of the sum of the similarity between $h_1$ and each $h_2 \in G_2$ to the number of hosts in $G_2$:

$$\textit{avg_similarity}(h_1, G_2) = \frac{\sum_{h_2 \in G_2} \textit{similarity}(h_1, h_2)}{|G_2|}$$

A partitioning $P$ of $I$ *respects avg_similarity* if for all $h_1 \in G_1$ and $G_2 \in P$, *avg_similarity*$(h_1, G_1) \geq$ *avg_similarity*$(h_1, G_2)$.

Respecting *similarity* or *avg_similarity* is not sufficient to generate a useful partitioning of $I$. After all, a partitioning that puts all the nodes in one group or one that puts each node in a separate group respects *similarity*. We therefore provide a parameter that can be used by network administrators to control how aggressive the algorithm is in partitioning $I$ into groups.

- Let $S_{min}$, the *similarity threshold*, be an integer greater $0$. A partitioning respects *similarity* and $S_{min}$ if it respects *similarity* and if, for $h_1$ and $h_2$ in $G$, *similarity*$(h_1, h_2) \geq S_{min}$.

- A partitioning $P$ of $I$ is said to be *maximal* with respect to *similarity* and $S_{min}$ if it respects *similarity* and $S_{min}$ and there does not exist another partitioning of $I$ that respects *similarity* and $S_{min}$ and has fewer groups. By adjusting $S_{min}$, one gets a maximal grouping with fewer groups in which the members of each group are more similar to each other.

### 3.1 Defining Similarity

We use connection behavior as a basis for host grouping, because that information is easily available by just monitoring the network. To group hosts, we need to define similarity in a way that captures the extent to which pairs of hosts establish connections to the same set of other hosts. We start by defining similarity between hosts as a function of the number of common hosts with which they communicate. Intuitively, hosts that share the same logical role communicate with similar sets of hosts.

A *connection* is a pair consisting of a source host address and a destination host address. The connection set of a host, $C(h)$, is the set, $\{a \mid a \in I$ and there is a connection between $h$ and $a\}$. If $h_1 \in C(h_2)$, then $h_2 \in C(h_1)$. We define the relation *neighbor*$(h_1, h_2)$ to be true if and only if $h_1 = h_2$ or $h_1 \in C(h_2)$. For later use, we extend the definition of *neighbor* to groups by defining *neighbor*$(G_1, G_2)$ to be true if and only if there exists a host $h_1 \in G_1$ that is a neighbor of another host $h_2 \in G_2$.

We can use the notion of a connection set to provide a simple definition of *similarity*:

$$\textit{similarity}(h_1, h_2) = |C(h_1) \cap C(h_2)| \quad (1)$$

That is to say that *similarity*$(h_1, h_2)$ is equal to the number of neighbors that $h_1$ and $h_2$ have in common.

We are now in a position to specify the requirements of a grouping algorithm. Given a set of hosts, $I$ and a similarity threshold, $S_{min}$, it must find a partitioning, $P$, of $I$ that is maximal with respect to *avg_similarity* and $S_{min}$, i.e.,

1. $P$ respects *avg_similarity*. This constraint guarantees that each host is within the group with which it has the strongest average similarity.

2. For all $h \in G$ and $G \in P$, *avg_similarity*$(h, G) \geq S_{min}$. This requirement guarantees that each host in a group is sufficiently closely related to every other host in the group, thus ensuring that groups are not too large.

3. There is no other partitioning $P$ of $I$ that meets the first two requirements and has a larger average group size. This guarantees that groups are not too small.

This specification is independent of the definition of *avg_similarity*. For some networks, such as the one represented in Figure 1, the above definition of *avg_similarity* yields excellent results. However, for others a slightly more complex definition works better. We present such a definition in Section 4.2.

## 4 Role Classification

The role classification problem is not difficult to solve in ideal situations, such as the network shown in Figure 1, in which two nodes that share the same logical role communicate with the identical set of machines. Clearly, such a situation does not reflect the connection patterns in typical enterprise networks. Three major challenges of the role classification problem are:

1. Two hosts that share the same logical role may communicate with drastically different sets of machines.

2. A host may potentially be classified into more than one role.

3. The grouping results that network administrators desire may vary from network to network and therefore the role classification algorithm must provide flexibility for them to control its mechanics so that meaningful grouping results can be achieved.

In a typical network setting for a technology company, each lab or test machine may be dedicated to a single engineer. Thus, each of these lab machines, despite sharing the same role, can have a connection pattern that is very different from the rest of the lab machines. To be able to correctly group such machines together, the grouping algorithm must take into account the potential roles of neighboring hosts rather than comparing the neighbor sets.

Furthermore, some hosts may potentially be classified into more than one role. For instance, there could exist a machine in the network in Figure 1 that communicates with both sets of machines with which many engineering machines and sales machines communicate respectively. In such cases, the connection patterns of hosts must be evaluated carefully to ensure that each host is grouped with other hosts with which it has the strongest similarity in connection habits.

The role classification problem is not trivial for the aforementioned reasons. Not only does the computation of the similarity measure matter, but the process of how nodes are grouped based on the similarity values among node pairs is also important.

The grouping algorithm consists of two phases: i) the group formation phase and ii) the group merging phase. The group formation phase identifies each group of hosts that have similar sets of neighbors using a simple similarity measure such as the one described in Section 3. The purpose of the group formation phase is two-fold: i) to efficiently identify various groups of hosts, each of which has drastically different overall connection patterns, and ii) to prepare for the second phase of the algorithm. The formation phase of the algorithm can efficiently find the desired partitioning for the example network in Figure 1 but may fail for many networks since it does not take into account the potential roles of neighboring hosts as explained earlier. In general, the group formation phase may generate a partitioning that contains more groups than desired.

The group merging phase decides whether groups, produced by the formation phase, can further be merged using a much more sophisticated similarity measure. This phase provides network administrators with fine-grained control over the merging process so that the grouping results reflect their intuition of the network structure.

### 4.1 Forming Groups

Group formation can be thought of as a graph theory problem. From the connection sets of $I$, one can generate a *neighborhood* graph, *nbh-graph*, where each node represents a host and each edge with weight $e$ represents that there are $e$ common neighbors between the hosts. Thus, a neighborhood graph captures the extent to which pairs of hosts communicate with the same set of other hosts. We use an undirected graph since almost all communication between hosts in the intranets is bi-directional. However, in certain situations, directionality may be used to improve the quality of the grouping results and we continue to investigate this issue.

One approach to the grouping problem is to treat it as a *k-clique* problem where *nbh-graph* is partitioned into cliques of size $k$ in which each edge in the clique has a weight greater than or equal to some constant $c$. Once a $k$-clique is identified, one assigns all the nodes in the $k$-clique to one group, since they all share at least $c$ common neighbors. This approach is problematic, because (i) the $k$-clique problem is NP-complete [25], and (ii) requiring that each host pair in a group has exactly $k$ common neighbors is too strong a requirement.

Another approach is to treat grouping as related to the problem of identifying bi-connected components (BCCs). A BCC is a connected component in which any two edges lie in a simple cycle. Thus, there exist at least two disjoint paths between any two nodes in a BCC. Unlike the $k$-clique problem, BCC can be solved in $O(N + E)$, where $N$ and $E$ are the number of nodes and edges in the graph respectively [9, 27]. Moreover, all nodes in the BCC need not be connected to each other directly. This approach is the one we use.

The group formation phase operates as follows:

1. Generate the connectivity graph, *conn-graph*, based on the observed connection patterns.

2. For $k = k_{max}$ down to 1, where $k_{max}$ is the maximum number of hosts with which a single host communicates:

   Repeat until no new groups can be assigned:

   (a) From *conn-graph*, build the k-neighborhood graph *k-nbh-graph*.

   (b) Remove group nodes (see 2d) from *k-nbh-graph*.

   (c) Generate all BCCs in *k-nbh-graph*.

   (d) For each BCC $B$, replace in *conn-graph* the nodes in $B$ by a new group node $G$ representing those nodes. Label $G$ by a pair $(ID_G, K_G)$, where $ID_G$ is a unique identifier and $K_G$ is $k$. ($K_G$ will be used later to compute the degree of similarity between groups.)

   (e) For each ungrouped host $h$, where $k < \alpha \times |C(h)|$ and $0 \leq \alpha \leq 1$, create a new group $G$ containing only $h$ as described in 2d.

The algorithm runs iteratively over *conn-graph* until no ungrouped node remains or $k = 0$. At each step multiple BCCs may be identified simultaneously and a single node could be a part of several BCCs indicating that it may share multiple roles. In this case, the node becomes a part of a BCC with the largest size. If more than one such BCC exists, we choose one randomly. By iterating over $k$ from high to low, the algorithm associates each host $h$ with other hosts with the strongest similarity.

In the grouping algorithm, the minimum number of nodes required to form a BCC is two. Technically, the minimum number of nodes to form a BCC is 3, since we do not allow duplicate edges between any two nodes. Nevertheless, we allow two isolated nodes connected by an edge to form a group.

Since a BCC is not a clique, some node pairs in the BCC may not have edges between them allowing node pairs that do not share at least $k$ common neighbors to be in the same group. However, any two nodes in a BCC have at least two disjoint paths along which two successive nodes share at least $k$ common neighbors. In other words, any two nodes in a group demonstrate in at least two different ways that they have strong similarity in connection habits, significantly reducing the possibility that they may serve different roles. This observation is a major reason why we believe BCCs are suitable for forming groups.

When a set of hosts is placed into a group, the nodes representing those hosts are removed from *conn-graph* and replaced by one node (called the *group node*) representing the entire group. There are edges connecting that group node to each node to which one of the hosts in the group was connected.

In some cases where a node may have connection patterns so different from any other nodes, the node should form a group by itself. Step 2e forms a new group with only $h$ in it if there exist no nodes that have the number of common neighbors greater than or equal to $\alpha \times C(h)$. We set $\alpha = 0.6$ and find that it works well with various networks.

Figure 2 illustrates the evolution of the grouping algorithm, in terms of *k-nbh-graph*, for the network depicted in Figure 1. The first group is formed when $k = M + N$, where $M$ is the number of hosts used by sales personnels and $N$ is the number of hosts used by engineers. For specificity, let us assume that $M = N = 3$. As shown in the picture, the *6-nbh-graph* con-

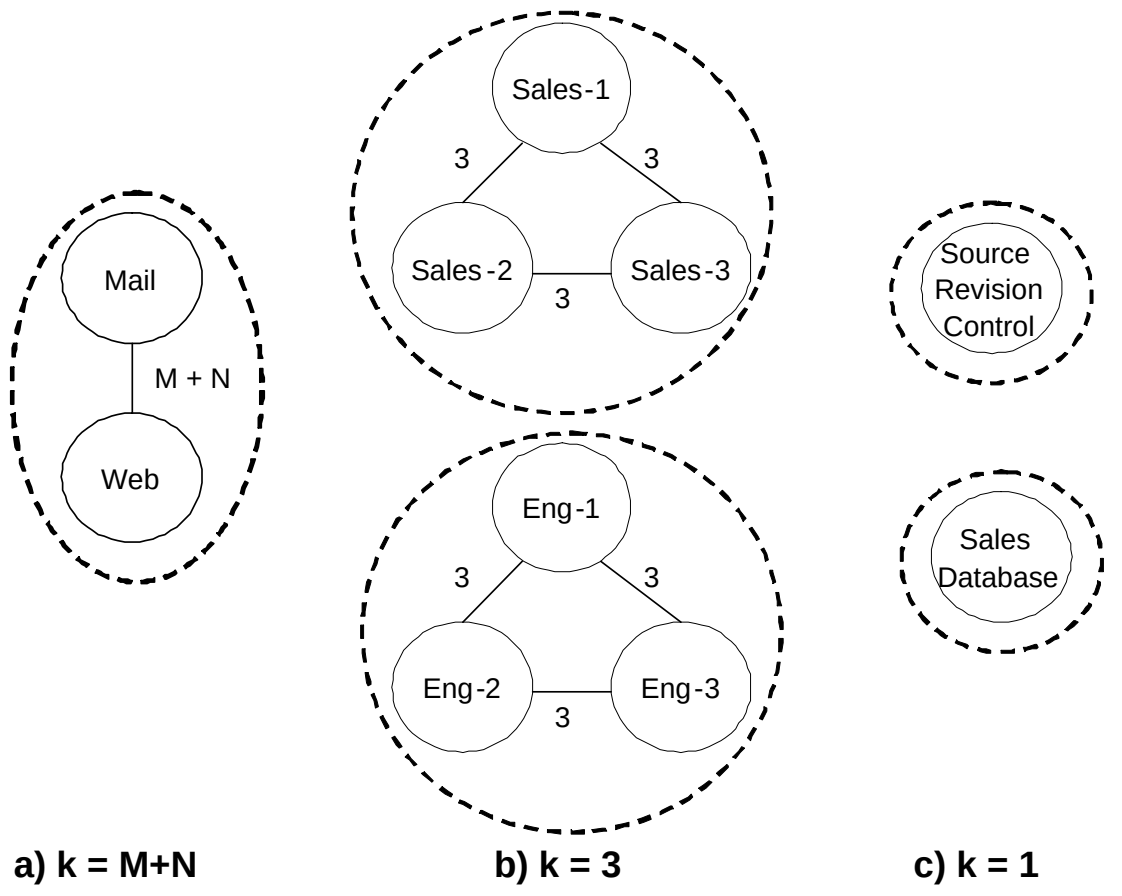

Figure 2. Evolution of the grouping algorithm at various $k$ values.

PROCEDURE MEETCONNECTIONREQ($G_1$, $G_2$) {
    a1 $\leftarrow \frac{\sum_{h1 \in G_1} C(h1)}{|G_1|}$
    a2 $\leftarrow \frac{\sum_{h2 \in G_2} C(h2)}{|G_2|}$
    return (a1 is within $\beta$ percent of a2)
}
PROCEDURE MEETSIMILARITYREQ($G_1$, $G_2$) {
    kmax $\leftarrow max(K_{G_1}, K_{G_2})$
    $s \leftarrow$ SIMILARITY($G_1$, $G_2$)
    if (kmax $\geq K^{hi}$ and $s \geq S_g^{hi}$)
        return true;
    else
        return (kmax $< K^{hi}$ and $s \geq S_g^{lo}$)
}
PROCEDURE SIMILARITY($G_1$, $G_2$) {
    $c1 \leftarrow \frac{\sum_{h \in C(G_1)} CP(h, G_1)}{|G_1|}$
    $c2 \leftarrow \frac{\sum_{h \in C(G_2)} CP(h, G_2)}{|G_2|}$
    For each common neighbor group $G'$ of $G_1$ and $G_2$
        $s \leftarrow s + min(\frac{CP(G', G_1)}{|C(G_1)|}, \frac{CP(G', G_2)}{|C(G_2)|})$
    $s \leftarrow \frac{1}{2} * (\frac{s}{c1} + \frac{s}{c2})$
    return $s * 100$
}

Figure 3. Pseudo-code for determining the similarity and connection requirements.

tains two hosts, *Mail* and *Web*, and the algorithm puts them in one group. When $k = 3$, the algorithm identifies two additional BCCs, one containing all the sales machines and the other, all the engineering machines. Finally, because of the bootstrap condition (see Step 2e), the algorithm creates two groups, one containing *Sales-Database* and the other, *SourceRevisionControl*, when $k = 1 < 0.6 \times M$.

## 4.2 Merging Groups

The aforementioned group formation algorithm that uses a simple definition of *similarity* tends to produce too many groups in many situations. Consider, for example, the network in Figure 1 modified so that *Sales-1* communicates with the *Mail* and *SalesDatabase* servers but not the *Web* server. The grouping algorithm in Section 4 will create two groups for the sales hosts, one that only contains *Sales-1* and one that contains the rest of the sales hosts. This might be appropriate, but it is probably not what a network administrator would want.

The group merging phase builds on the results generated by the group formation phase. It merges groups that are similar in connection habits in a way that allows users to control the process so that more meaningful results can be achieved.

During the grouping phase, we merge two groups if they meet the following two requirements:

**Similarity requirement.** The similarity measure between the two groups exceeds a user-specified threshold.

**Connection requirement.** The average number of connections of each group is comparable.

The algorithm repeatedly merges two groups that meet the two requirements and have the highest similarity measure until no groups can be merged. The $K$ value of a newly merged group is set to the minimum number of connections a host in the group has.

Figure 3 depicts the pseudo-code for determining the average connection requirement and the similarity requirement. The procedure MEETCONNECTIONREQ decides whether the two groups, $G_1$ and $G_2$, meet the connection requirement. This requirement keeps a group with a large number of connections from merging with another group with a much smaller number of connections.

The procedure MEETSIMILARITYREQ determines whether the two groups meet the similarity requirement. $S^{hi}$ and $S^{lo}$, $S^{hi} > S^{lo}$, are similarity thresholds that can be set by the user to control the merging process. Which threshold is used depends upon whether $max(K_{G_1}, K_{G_2}) \geq K^{hi}$, where $K^{hi}$ is a constant intended to define whether a $K_G$ value is "high." The similarity threshold for merging groups is higher for groups with a high $K_G$ value, those groups whose member hosts share a high number of common neighbors. This is because merging two groups can change the relationships between other groups in a way that induces additional undesirable merges.

Again, consider the groups in the example network illustrated in Figure 1. Notice that if $N$ is large, the similarity measure between the *SalesDatabase* group and the *Mail* and *Web* group will be large. Similarly, for large $M$, the *SourceRevisionControl* group will be highly similar to the *Mail* and *Web* group. If all three groups were to merge, it would effectively cause the sales group and the engineering group to merge, resulting in a partitioning with two groups: one containing all the servers, and one containing all other hosts. In most situations, this grouping would be undesirable since the network administrators lose the important separation between the *Sales* machines and the *Eng* machines. For these reasons, groups with high $K_G$ values are required to have a higher similarity measure to merge. We discuss how best to choose the constants in Section 6.

SIMILARITY computes the similarity $s$ (between 0 and 100) of connection patterns between two groups. $CP(x_1, x_2)$ returns the total number of connections between $x_1$ and $x_2$, where $x_i$ could either be a host or a group. Two groups are considered similar if they have many common neighboring groups and similar average numbers of connections. For example, if the set of neighbors of $G_1$ is a subset of the set of neighbors of $G_2$, it increases the desirability of merging these two groups. However, if the average numbers of connections of $G_1$ and $G_2$ are quite different, the desirability of merging them is lessened.

## 5 Role Correlation

Over time, connection habits may evolve as new servers and employees are added while some existing ones leave. Sometimes hosts may behave erratically as a result of being victims or villains of denial of service (DOS) attacks. Due to any of these behaviors, the grouping algorithm may produce a different set of groups than the one produced by the algorithm a few days ago. As explained in Section 4, the grouping algorithm assigns an integer ID to each group of hosts that it identifies. There is no guarantee that the sets of IDs produced by two runs of the grouping algorithm will have any correlation between them. This situation is clearly undesirable to the users who may want to associate logical names and policy settings to the group IDs and preserve these group specific data throughout the executions of the grouping algorithm at various times.

In this section, we describe in detail the group correlation algorithm that takes the two sets of results produced by the grouping algorithm and correlates the IDs of one set with that of the other so that the two groups, one in each set of resulting groups, will have the same ID if and only if the machines in both groups are highly likely to share the same logical role.

### 5.1 Challenges

For the rest of this section, we assume that there exists a unique host identifier that never changes. We note that the IP address may not be a good use when Dynamic Host Control Protocol (DHCP) is used since a host's IP address may change over time. For smaller networks, a simple solution such as using DNS names as unique identifiers and dynamically updating the changes of IP addresses may be sufficient [26]. This problem of assigning a unique identifier to each host within enterprise networks is beyond the scope of this paper.

The connection habits of a host may change as a result of the following events: i) new host arrivals, ii) existing host removals, and iii) role changes by existing hosts. Due to a combination of these events, some existing hosts may communicate with different sets of hosts and thus the results of the grouping algorithm before and after these events may be different as: i) new groups are formed, ii) existing groups are deleted, iii) the member compositions of some groups change, and iv) the connection sets of some groups change. The changes affect not only the hosts directly involved in the aforementioned events but also to other hosts whose connection habits have not changed in a logical sense.

Hypothetically, if we know the exact sequence of every single change event that happened between two executions of the role classification algorithm, the results of the first execution could be incrementally updated to achieve the new results. Having such a change log, although not impossible, can complicate the network data gathering process. More importantly, a detailed change log cannot always lead to correct ID correlations.

Consider the example network in Figure 1. Assume that *Sales-1* and *Eng-1* switch roles as a result of personnels switching jobs or changing machines. *Sales-1* now communicates with *SourceRevisionControl* whereas *Eng-1* communicates with *SalesDatabase*. From the change log, it would seem that the connection sets of both *SourceRevisionControl* and *SalesDatabase* change whereas in reality, their logical roles never changed. The difficulty here is in distinguishing which changes in connection patterns are the primary causes that result in differences in group formations between two executions of the grouping algorithm. Furthermore, there may also be natural changes in connection patterns of many nodes. For instance, an existing server machine may be replaced by two new machines that do load sharing among client machines. The logical roles of the client machines have not changed but their observed connection patterns have. The rest of this section describes the role correlation algorithm that does not rely on the change log but rather uses the same set of information (i.e. only connection sets) made available to the grouping algorithm.

## 5.2 The Role Correlation Algorithm

The correlation algorithm operates by comparing the results of two executions of the grouping algorithms. Let $P_{t-1}$ and $P_t$ be the group sets generated by the grouping algorithm at time $t-1$ and $t$ respectively. The correlation algorithm updates the ID set of $P_t$, so that $ID_{G_t} = ID_{G_{t-1}}$, where $G_t \in P_t$ and $G_{t-1} \in P_{t-1}$, if and only if $G_t$ and $G_{t-1}$ considered to represent the same logical role. More specifically, the connection patterns of the members of $G_t$ and those of $G_{t-1}$ are very similar. The groups correlation algorithm correlates the $ID_{G_t}$ and $ID_{G_{t-1}}$ in a meaningful manner and thus allow applications to preserve data specific to a particular group.

The role correlation algorithm:

- Isolates the primary events, such as node arrivals and removals, that directly affects the connection habits of groups,
- Identifies nodes that have not changed their neighbors,
- Heuristically computes the *time-varying* similarity between the connection habits of two groups formed at times $t$ and $t-1$, and assigns $ID_{G_t} = ID_{G_{t-1}}$ if and only if the role of hosts (in terms on their connection patterns) in $G_{t-1}$ can be considered identical to the role of hosts in $G_t$.

First, the correlation algorithm eliminates the differences between the two host sets, $I_t$ and $I_{t-1}$, so that it can compare the connection patterns meaningfully. The algorithm computes the set of nodes that existed at time $t-1$ but have been removed in time $t$ ($I_{t-1} \setminus I_t$), and the set of nodes that only appear at time $t$ ($I_t \setminus I_{t-1}$). All new nodes are removed from $I_t$ and deleted nodes are removed from $I_{t-1}$. Thus, the changes in connection set of each host is only as a direct result of changing connection patterns between the host and its neighbors (which existed at time $t$).

Second, the algorithm heuristically identifies the set, $H_{same}$, of nodes that are very like to play the same logical roles from $t-1$ to $t$. We say that the two nodes $h_t$ and $h_{t-1}$ are highly likely to be the same machine (i.e. it hasn't changed its logical role) if they have the identical connection sets. Specifically, $H_{same} = \{h_t | \exists h_{t-1} \in I_{t-1}, C(h_t) = C(h_{t-1})\}$. We will explain shortly how we use the fact that a host $h \in H_{same}$ to our advantage in computing the time varying similarity measure.

The role correlation algorithm will determine whether the two groups $G_t$ and $G_{t-1}$ are the same group by heuristically computing the time-varying similarity measure and comparing against the pre-defined threshold. The group correlation algorithm operates as follows:

1. For each group $G_t$, identify $G_t$ and $G_{t-1}$ as the same group if i) $G_{t-1}$ has the strongest time-varying similarity with $G_t$, among all the groups in $P_{t-1}$ and ii) the average number of connections is at least within $T^{hi}$ percent of the average number of connections of $G_{t-1}$.

2. For each group pair$(G_t, G_{t-1})$ that remain uncorrelated, decide whether $G_t$ and $G_{t-1}$ represent the same logical group based on how similar the connection patterns between $G_t$ and its neighbor groups are to those between $G_{t-1}$ and its neighbor groups.

Step 1 decides whether the two groups $G_t$ and $G_{t-1}$ are identical based on the time varying similarity measure. As in Section 4.2, we compute the similarity measure based on the average number of connections between the groups and their common neighbors. However, finding the common neighbor set between $G_t$ and $G_{t-1}$ is not trivial. This is because we cannot simply assume that a neighbor $h_t \in C(G_t)$ and a neighbor $h_{t-1} \in C(G_{t-1})$ are the same host even if they have the same host identifier. We use the following techniques to identify the common neighbor set:

- If a neighbor $h_t$ of $G_t$ shares the same host identifier with the neighbor $h_{t-1}$ of $G_{t-1}$ and both have been considered highly likely to be the same host (i.e. $h_t, h_{t-1} \in H_{same}$), we assume $h_t$ is the neighbor to $G_t$ in the same way as $h_{t-1}$ is to $G_{t-1}$.

- For each neighbor pairs $(h_t, h_{t-1})$ that are not considered as highly likely to be the same host, we assume $h_t$ is the neighbor to $G_t$ in the same way as $h_{t-1}$ is to $G_{t-1}$ if and only if the following condition is true. The connection set size of $h_{t-1}$ is within $T^{hi}$ percent of that of $h_t$ and no other neighbor of $G_{t-1}$ has the connection set size closer to that of $h_t$.

The algorithm then computes the time-varying similarity measure between each neighbor pair $(h_t, h_{t-1})$, which meets the aforementioned requirements, as the minimum of the average number of connections between $h_t$ and $G_t$ and between $h_{t-1}$ and $G_{t-1}$. If the sum of the similarity measures for all common neighbor pairs within the bounds of the specified thresholds , the algorithm declares that groups $G_t$ and $G_{t-1}$ mean the same group.

# 6 Results

In this section, we evaluate the performance of the algorithms using traces gathered over a day at two corporate networks. We show that the algorithms operate well for both networks and examine the effects of user-defined thresholds on the results of the role classification algorithm.

We call the two test networks *Mazu* and *BigCompany*. Mazu is part of the corporate network at Mazu Networks, Inc., in Cambridge, MA. It consists of 110 hosts, including engineering workstations, several servers, and laptops. Mazu develops various software products in the area of network security and monitoring. The BigCompany network consists of 3638 hosts, including workstations, servers, and many IP phones. For privacy reasons, BigCompany must remain anonymous.

## 6.1 Effectiveness of the Grouping Algorithm

We evaluate the effectiveness of the role classification algorithm by comparing the groups formed by the algorithm against the logical roles that hosts play as determined by knowledgeable network administrators. For all the experiments, unless otherwise noted, we set user-defined thresholds, $S^{hi} = 80, S^{lo} = 55$, and $K^{hi} = 7$. We examine how these thresholds affect the results in later sections.

Figure 4 shows some of the groups formed by the role classification algorithm running on the Mazu data and configured with the default parameters. Each circle in the figure represents a group and lists its members and its connections with other groups. Where possible, we have indicated the logical role of each host, which we obtained by asking the Mazu network administrator. (Of course, this logical information was not used in constructing the grouping.)

Observe that the role classification algorithm placed almost all engineering (*eng*) machines in a single group, 85. Also note that the number of connections of an engineering host varies from 4 to 9. Similarly, most machines used by sales, management (*admin*) and operations (*ops*) were placed in a single group, 87. The largest group, 80, contains new machines and test machines in the lab.

However, four hosts that are identified as engineering machines are placed in group 87 rather than group 85. The reason is that these machines do not communicate with a set of hosts that engineering machines in group 90 communicate with. As shown in Figure 4, each engineering machine in group 90 has, on average, one connection with group 10, which consists of a Unix mail server, and one connection with group 6, which consists of a source revision control server (not shown in the figure). On the other hand, almost every sales host in group 87 communicates with both the Microsoft Exchange server and the NT sever from group 71, but not with the Unix mail server nor the source revision control server. In fact, there are just two connections between group 87 and each of groups 6 and 10. The four engineering hosts in 87 had connection patterns very similar to those of sales hosts, so they were grouped accordingly. Most probably, these machines are used by engineering managers who do not perform engineering-related tasks such as coding, and use the Exchange mail server instead of the Unix mail server.

If we have the perfect knowledge of the logical structure of the network, we can use that knowledge to quantify the resulting quality of the groups produced by grouping algorithms. One simple yet effective metric used in the cluster validation literature [16, 12] is *Rand Statistic*, which is based on testing whether a pair of objects belongs to the same group as decided by the grouping algorithm and according to our knowledge. Let $P$ and $P^*$ be the partitionings of hosts produced by a grouping algorithm and based on our knowledge respectively. Let $SS$, $SD$, $DS$ and $DD$ be the numbers of host pairs that belong to the same group in both $P^*$ and $P$, to the same group in $P^*$ and to different groups in $P$, to different groups in $P^*$ and to the same group in $P$, and to different groups in both $P^*$ and $P$ respectively. $SD$ and $DS$ are indicative of how different $P$ is from $P^*$. Rand Statistic $R = \frac{SS+DD}{SS+SD+DS+DD}$ is between 0 and 1. The higher the value, the more similar $P$ and $P^*$ are.

For the Mazu network, we were able to ascertain the logical roles of all except 8 hosts. We worked closely with the Mazu network administrator to obtain $P^*$, the ideal partitioning of hosts. We find that the partitioning produced by the grouping algorithm (with default parameters) achieves $SS = 452, SD = 710, DS = 133, DD = 3856$ and $R = 0.8363$. This shows that the results of the grouping algorithm reflect to a high degree our intuitive notion of the underlying structure of the network. We note that the reason for having a relatively high $SD$ is because the algorithm identifies subsets of hosts within large groups as separate groups. For instance, the grouping algorithm produces a few different groups, each containing a single *eng* machine instead of putting them in group 80 (not shown in the figure). This is because those *eng* machines have the total number of connections far greater than the average number of connections that a host in group 80 does. Such distinction may prove useful in certain situations.

Table 1 lists the five largest groups produced by running the grouping algorithm on the BigCompany network. Again, we relied on information generated by the network administrator to help us understand whether the groupings generated by the algorithm matched the log-ical structure of the network. Group 1020 consists of desktops whose IP addresses are managed by the DHCP server. Almost every machine in group 1020 communicates with approximately 85% of the machines in group 1075, and vice-versa. This pattern suggests that it was appropriate for the grouping algorithm to combine the machines in group 1075, which use static IPs, into a group. Most machines in both groups run Microsoft Windows. The high number of connections between the groups is due to Windows file sharing, which uses the NetBIOS network protocol. File sharing creates a large number of connections between the hosts in the two groups, even though in both groups there is little intra-group communication. We continue to investigate this interesting relationship between the two groups. It is striking, and further proof of the need for better analysis tools, that the network administrators we have talked to themselves don't know why the groups are partitioned in this way.

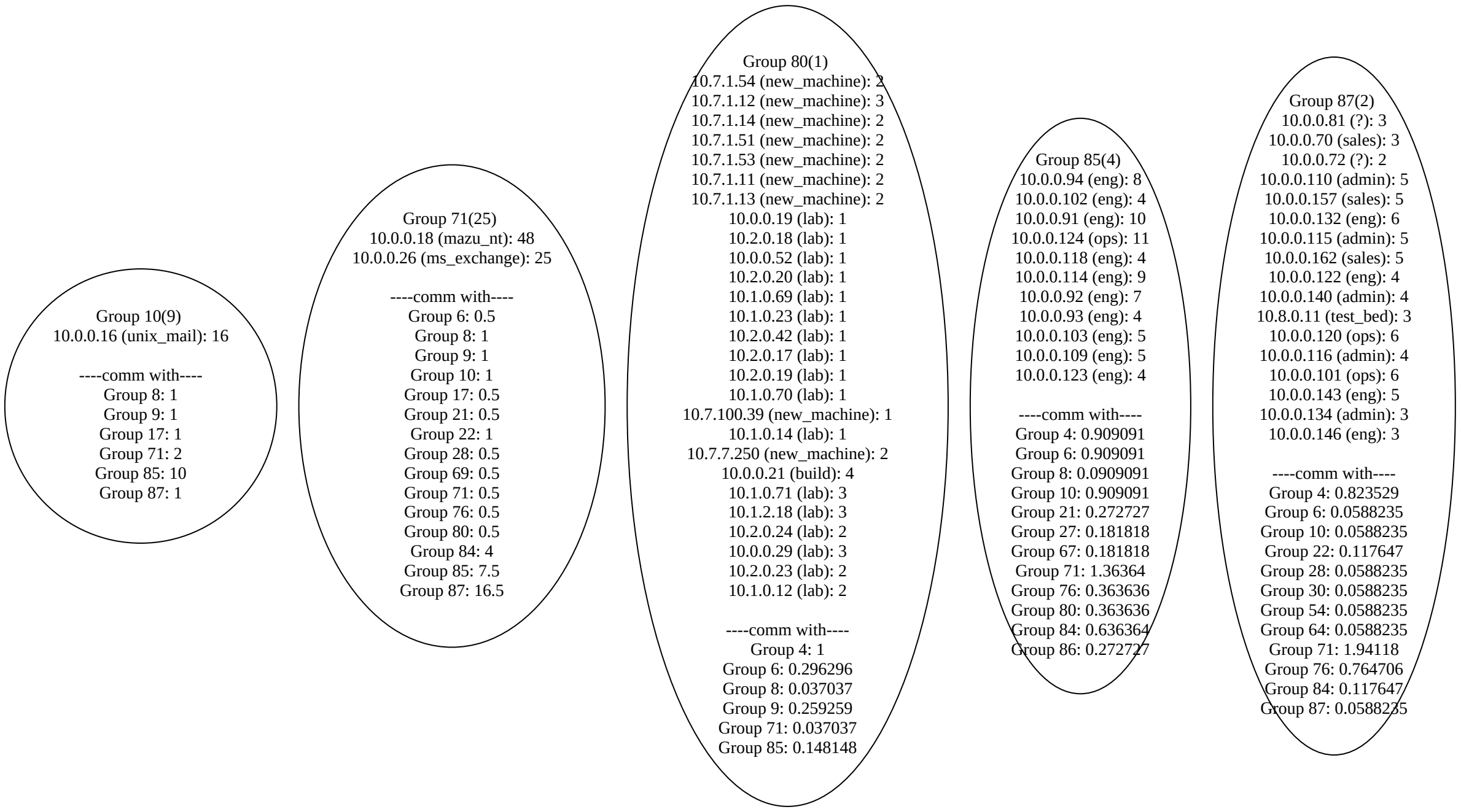

Figure 4. Grouping results based on data gathered over one day at Mazu. The number in parentheses next to the group ID is the group's $K_G$. The number next to each host is a count of the host's connections. Each line after "comm with" denotes a neighbor group and the average number of connections between the group and that neighbor.

The grouping algorithm also correctly classifies all IP phones into one group, 1092. Group 1138 consists of web servers and other servers that desktops in group 1020 regularly communicate with. Group 1043 is the largest group, with 1519 members. Most machines in this group have a single connection (hence the role name *idle*), and that is to a host that opens connections to about 1,600 machines. With our help, BigCompany is currently investigating why this host scans about 45% of all the machines on the network. This example is another good use of how the role classification algorithm can be applied to understand networks and detect anomalous behavior.

Table 2 summarizes the grouping results of the two networks. Observe that the number of groups in the BigCompany network is 26 times smaller than the number of hosts. Unfortunately, we cannot use Rand Statistic to quantify the quality of the groups produced by the grouping algorithm since we don't have the perfect knowledge of the logical roles of each machine in the BigCompany network. Nevertheless, the network administrators at BigCompany report that they find them both useful and consistent with their intuitions about their networks. We are also in the process of analyzing a larger network owned by HugeCompany that consists of 49,041 hosts.

### 6.2 Effectiveness of the Correlation Algorithm

This subsection shows that for a specific scenario, the role correlation algorithm associates new groups with existing ones in an appropriate way. Figure 5 lists the scenario we investigate. In the Mazu network, we swapped the roles of *unix_mail* and *ms_exchange* by switching their IP addresses. We also replaced the old NT server, called *mazu_nt* (10.0.0.18), with a new server (10.0.1.18). Finally, we removed an old *admin* machine (10.0.0.110) and brought in a new *eng* machine (10.0.0.200). Although the specific scenario is just one of many possible ones, it includes the types of changes that could happen in a real network.

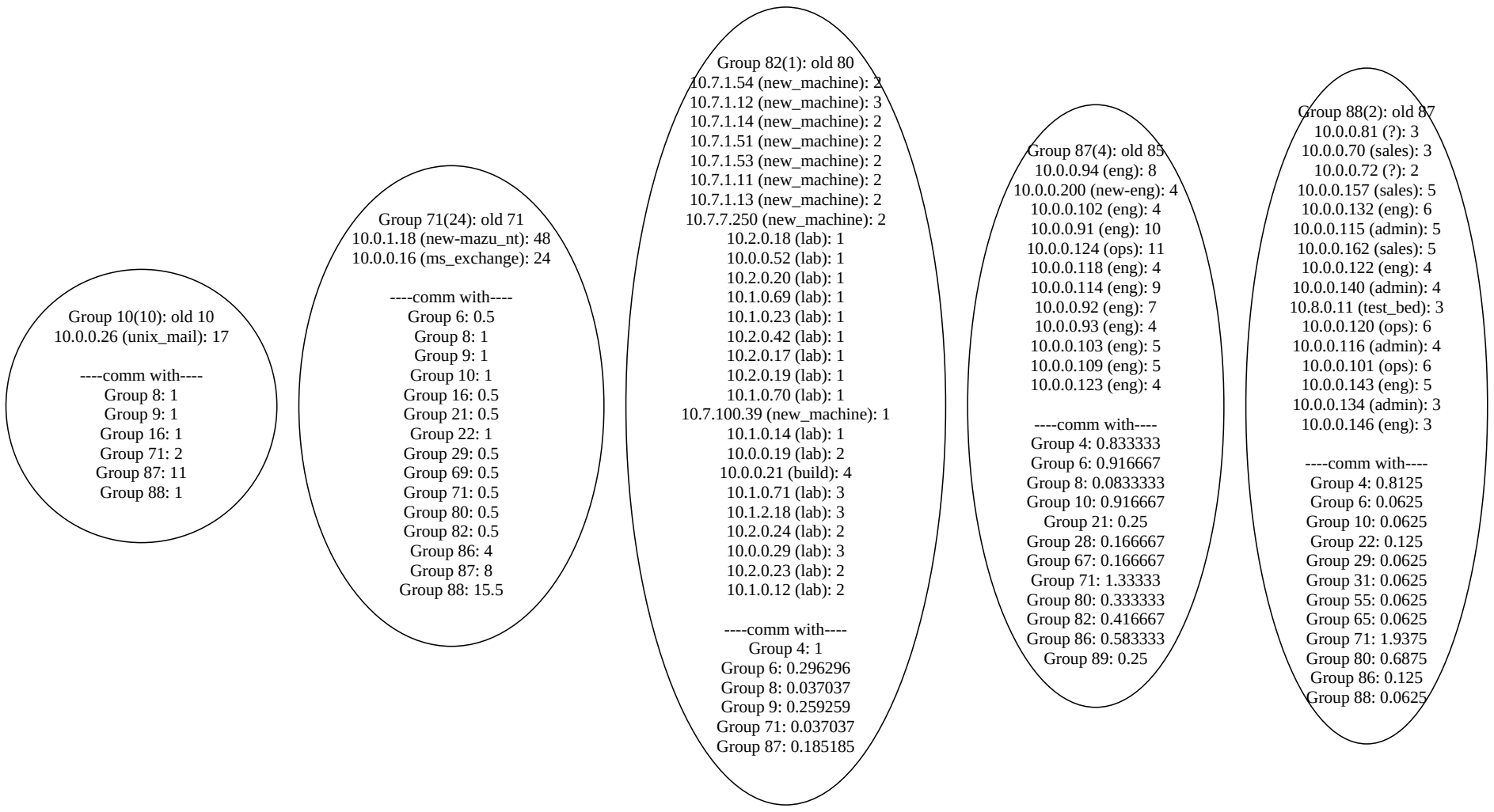

| Host | Previous role | New role |
|---|---|---|
| 10.0.0.16 | unix_mail | ms_exchange |
| 10.0.0.26 | ms_exchange | unix_mail |
| 10.0.0.18 | mazu_nt | - |
| 10.0.1.18 | - | mazu_nt |
| 10.0.0.200 | - | eng |
| 10.0.0.110 | admin | - |

Figure 5. The grouping results on the Mazu network with several changes (see table) to the connection patterns. The number next to "old" represents the ID of the correlated group shown in Figure 4.

| Group ID | Members | Logical Role |
|---|---|---|
| 1043 | 1490 | Idle |
| 1020 | 158 | DHCP-Desktops |
| 1138 | 396 | Servers |
| 1092 | 167 | IP-Phones |
| 1075 | 156 | StaticIP-Desktops |

Table 1. The five largest groups classified in BigCompany network that consists of $3638$ hosts. Logical role is identified by knowledgeable network administrators at BigCompany.

The modified connection patterns were used as inputs to the role classification algorithm. The role correlation algorithm then correlated the new grouping results with the original results. Every group in the new results is correlated with an old group. Figure 5 depicts the four groups that are affected by the changes. Observe how the member compositions of these four groups change from the ones in Figure 4. Both *unix_mail* and *ms_exchange* are correctly identified in the same fashion as in Figure 4 despite their role reversal. The new NT server (*new-nt_server*) appropriately takes the place of the old one. Similarly, a new *eng* host is grouped with other *eng* machines. Despite various changes to the connection patterns, the role correlation algorithm was able to correctly associate each new group with an existing one. We continue to investigate the limits of the role correlation algorithm under rigorous changes in connection patterns.

### 6.3 Configuration

The algorithms use two internal constants that we believe are not sensitive to particular network connection patterns. The group formation phase of the role classification algorithm (see Section 4.1) requires a constant $0 \leq \alpha \leq 1$ to keep a host $h$ from forming groups with other hosts that have less than a fraction $\alpha$ of the number of connections that $h$ has. The group merging phase keeps the two groups from merging if the average number of connections of a group is not within $0 \leq \beta \leq 1$ of the other's (see Figure 3).

We set $\alpha = 0.6$ and $\beta = 0.5$. Our experiments with both Mazu and BigCompany networks indicate that the default values work well on at least two rather different networks. We believe that, in general, it will not be necessary to adjust these constants. Nevertheless, we plan to expose these parameters to network administra-tors so that they can adjust them along with the similarity thresholds to achieve grouping results that most reflect their intuition of the network structure.

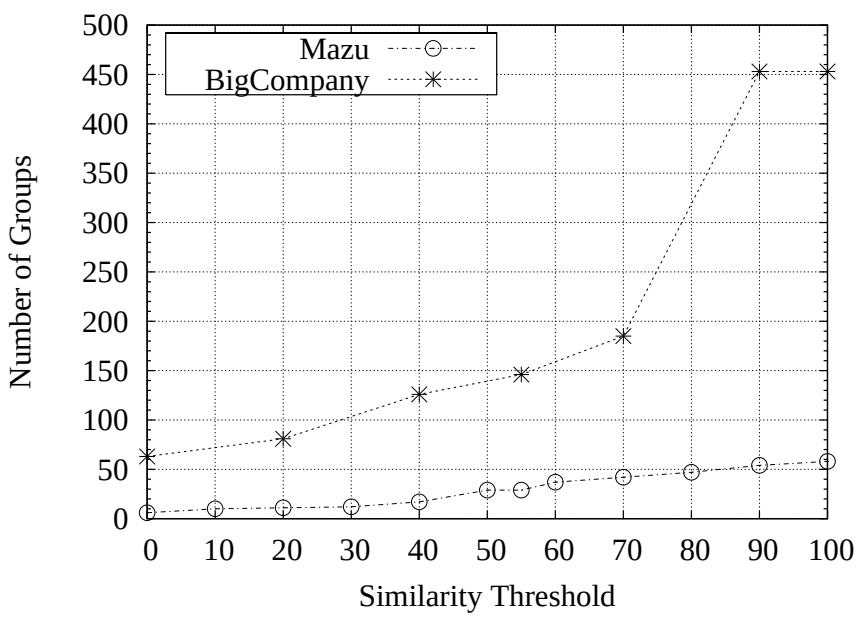

Figure 6. Number of Groups vs. $S^{lo}$

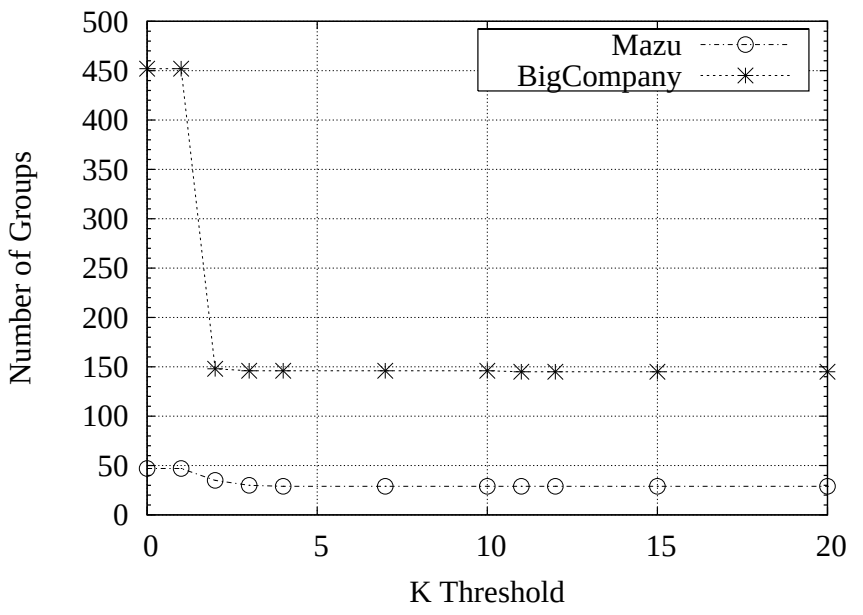

Figure 7. Number of Groups vs. $K^{hi}$

### 6.4 Effects of Similarity Thresholds

In this subsection, we examine how the choice of the user-defined thresholds, $S^{lo}$, $S^{hi}$, and $K^{hi}$, affect the number of groups formed by the role classification algorithm. Recall that the two groups are merged if and only if their similarity measure is $\geq S^{lo}$. Furthermore, if the maximum $K_G$ associated with the groups is $\geq K^{hi}$, they are not merged unless their similarity measure is $\geq S^{hi}$. We require that $0 \leq S^{lo} < S^{hi} \leq 100$.

Figure 6 illustrates how $S^{lo}$ affects the total number of groups formed for both Mazu and BigCompany networks. The number of groups increases with $S^{lo}$. Again, a large $S^{lo}$ value keeps more groups from merging and as a result, the total number of groups remains large.

The number of groups may not increase smoothly with the increase in $S^{lo}$. For instance, there is steeper incline (knee) in the number of groups of BigCompany network when $S^{lo}$ is increased from 70 to 90. The reason is that the increase in $S^{lo}$ causes some groups with high numbers of connections to split, since they no longer meet the stronger similarity requirement to merge. This in turn causes several neighboring groups to split. The extent to which such splits occur varies from network to network. A knee in the curve indicates that the algorithm can expose the logical structure of the network in two significantly different manners. Consider again the network in Figure 1. If $S^{lo}$ is too low, *Mail*, *Web*, *SalesDatabase*, and *SourceRevisionControl* will all be placed in one group, whereas all sales and engineering machines will be placed in another. In some cases, such grouping might be more appropriate than the one achieved in Figure 1. Network administrators should compare the grouping results on both sides of the knee and decide which one better suits their needs.

Our experiments show that as long as $S^{hi} \geq 80$, changes to $S^{hi}$ hardly affect the grouping results. Therefore, we suggest that $S^{hi}$ be fixed.

On the other hand, the choice of $K^{hi}$ has a significant impact and should probably vary from network to network. If $K^{hi}$ is set to the maximum number of connections that any host has, the similarity measure between hosts is only compared against $S^{lo}$. If $K^{hi} = 0$, the similarity measure is only compared against $S^{hi}$. Ideally, $K^{hi}$ should be set at a value that partitions the hosts in the network into two groups, one containing all server-like machines, and one containing all others.

Figure 7 shows how $K^{hi}$ affects the number of groups formed. For any two data points with the same number of groups, the grouping results are identical. Clearly, the grouping results do not change for the Mazu network when $K^{hi} \geq 4$. Similarly, the grouping results hardly change for the BigCompany network when $K^{hi} \geq 3$. This implies that it is not too difficult to find an appropriate $K^{hi}$ for a particular network. By default, we set $K^{hi} = 7$ and believe that this value will be suitable for most networks. Nevertheless, we are currently working on automatically setting $K^{hi}$.

### 6.5 Run Time

Table 2 shows the time taken to run the role classification algorithm on the Mazu and BigCompany networks. We performed our experiments on a Linux machine equipped with a 2GHz Intel Xeon processor and 4GB of memory. The run time achieved by the algorithms grows quadratically with the number of nodes and is acceptable for use in commercial enterprise network monitoring and analysis tools. We continue to further improve the performance of the algorithms.

## 7 Related Work

The work described in this paper was implemented in part using Click [21], a modular router system that makes it easy to build efficient packet processing devices on commodity PC hardware. The grouping al-gorithm only requires information about connections among hosts, so it can obtain data from a variety of sources, from summary formats like RMON [28] and Netflow [6] to packet-level sniffers like tcpdump [18].

| Network | Hosts | Groups | Run time(s) |
|---|---|---|---|
| Mazu | 110 | 25 | 0.069 |
| BigCompany | 3638 | 137 | 63 |
| HugeCompany | 49041 | 1374 | 2101 |

Table 2. The summarized grouping results for Mazu and BigCompany networks.

Part of the role classification algorithm can be viewed as a data clustering algorithm. Data clustering has been an active area of research for a few decades [14, 19] and is known to be a difficult problem combinatorially. The techniques used to cluster data vary widely according to the assumptions, and contexts specific to application domains and many existing techniques are specifically developed for pattern recognition and image analysis. In general, a data clustering algorithm attempts to cluster data points or *patterns*, each of which is represented by a vector of real numbers. Patterns that are similar to each other are clustered together. The most popular metric for similarity measure is the Euclidean distance. One well-known clustering technique is the *hierarchical agglomerative clustering* technique. The idea is to merge clusters based on the pair-wise similarity measure of patterns. The merging process is stopped according to some predefined similarity thresholds. In this aspect, the group merging phase of the role classification algorithm can be classified as a hierarchical agglomerative clustering technique.

The main reason why traditional data clustering algorithms cannot be easily extended for our application domain is because it is difficult to represent the connection pattern of each host with a vector of numbers in such a way that the widely used Euclidean distance to measure the similarity between two connection patterns makes sense. Furthermore, we can leverage the communication patterns found in typical enterprise networks, such as client-server communications, to achieve more meaningful grouping results. We also note that traditional data clustering techniques do not deal with temporal correlation of clusters as the role correlation algorithm does.

The role classification algorithm is applicable to network intrusion detection. For example, grouping information provides context that can be used by intrusion detection systems [10, 22] (IDS) to help determine how unusual (and hence potentially suspicious) a certain network behavior is (see Section 2).

As explained in Section 2, role grouping is well-suited to improving network monitoring and policy management. An entire industry [8, 15, 17] caters to enterprises' network management needs, and much literature is devoted to network monitoring, traffic reporting, and performance measurement [13, 20, 23, 24]. All this work differs significantly from ours. The commercial network management systems are primarily integration and alerting tools, intended to provide operators with a unified view of disparate devices on the network. They serve as conduits for the raw data, but do not extract higher-level semantics such as role relationships. Academic work has focused on network monitoring and techniques for performance measurement, but again, the interpretation of data is generally left to humans.

Another tool that can help operators understand their networks is network visualization [1, 5, 11]. Visualization focuses on graphic design and automated layout algorithms to help users digest the vast amount of data generated by network monitoring tools. Unlike the grouping algorithm, these techniques have no notion of the logical structure of the network. However, they can complement grouping, exposing grouping information to the user and using grouping information to make better decisions about visual layout.

## 8 Summary

This paper has presented two practical algorithms (grouping and correlation) that group hosts on an enterprise network into roles according to their observed connection patterns. The first algorithm partitions hosts on the network into groups based on connection data. The second algorithm meaningfully correlates the results obtained by running the first algorithm at different times, taking into account the evolution of connection patterns over time.

To our knowledge, the problem of automatically grouping and classifying hosts based on their behavior on the network has not been addressed before. This paper formulates the problem by presenting an abstract model in addition to the concrete algorithm specifications. The general framework we have developed accommodates other classification algorithms in addition to the ones we have described.

Grouping hosts according to their connection habits exposes the logical structure of the network, and can serve to improve understanding of the network and to simplify a variety of network management tasks. It can also improve the accuracy of automated tools, such as systems for network monitoring and intrusion detection.

Experience with the algorithms on two corporate networks, one with about 100 hosts and one with over 3600 hosts, indicates that they work well. They are easy to tune, and produce results that are meaningful and consistent with the intuition of experienced network administrators. Importantly, our experience on the corporate networks has shown that automated classification algorithms such as these can play an important role in assisting network administrators. The algorithms are also fairly efficient, and their performance remains practical even for networks with several thousand hosts.

Much work remains to be done. We plan to continue improving the performance of the algorithm. The ideal solution should be better than quadratic time complexity, since that could eventually be the limiting factor on very large networks. We will also explore other definitions of host similarity for grouping. For instance, one could consider incorporating services (such as TCP or UDP port information) or protocols into the definition of a connection, so that a web server would not be grouped with a mail server. In addition, we have yet to explore many of the applications of automatically-derived grouping information, which include network management, provisioning, security, and perhaps others.